\documentclass[10pt,twocolumn,twoside]{IEEEtran}
\usepackage{balance}
\usepackage{amsmath}
\usepackage{amssymb}
\usepackage{dsfont}
\usepackage{graphicx}
\usepackage{theorem}
\usepackage{cite}
\usepackage{pstricks}
\usepackage{epsfig}
\usepackage{siunitx}
\usepackage{booktabs}
\usepackage{pst-grad} % For gradients
\usepackage[normalem]{ulem}   %Nima added this, to be removed later
\usepackage{bbm}
\usepackage{float}
\usepackage{footnote}
\DeclareMathOperator{\im}{im}

{\theorembodyfont{\itshape}\newtheorem{theorem}{Theorem}}
{\theorembodyfont{\itshape}}
{\theorembodyfont{\itshape}}
{\theorembodyfont{\itshape}\newtheorem{corollary}{Corollary}}
{\theorembodyfont{\upshape}\newtheorem{definition}{Definition}}
{\theorembodyfont{\itshape}}
{\theorembodyfont{\upshape}\newtheorem{remark}{Remark}}
{\theorembodyfont{\upshape}}
{\theorembodyfont{\upshape}\newtheorem{example}{Example}}
{\theorembodyfont{\upshape}\newtheorem{assumption}{Assumption}}

\newcommand{\1}{\mathds{1}}
\newcommand{\R}{\mathbb{R}}
\newcommand{\w}{\omega}

\newcommand{\pu}{\mathrm{pu}}

\newcommand{\half}{\frac{1}{2}}
\newcommand{\la}{\mathcal{L}}
\newcommand{\Y}{\mathcal{Y}}

\newcommand{\I}{\mathcal{I}}
\newcommand{\U}{\mathcal{U}}
\newcommand{\diag}{\mathop{\mathrm{diag}}\nolimits}
\newcommand{\mH}{\mathrm{mH}}
\newcommand{\redu}{\mathrm{red}}

\usepackage[prependcaption,colorinlistoftodos]{todonotes}
\newcommand{\ER}{\hspace*{\fill}$\square$}
\begin{document}
\title{Output Impedance Diffusion into \\Lossy Power Lines}

\author{Pooya~Monshizadeh,\; Nima Monshizadeh,\; Claudio De Persis,\; and Arjan van der Schaft% 
\thanks{Pooya Monshizadeh and Arjan van der Schaft are with the Bernoulli Institute for Mathematics, Computer Science, and Artificial Intelligence, University of Groningen, 9700 AK, the Netherlands,
        {\tt\small p.monshizadeh@rug.nl, a.j.van.der.schaft@rug.nl}}%
\thanks{Nima Monshizadeh and Claudio De Persis are with the Engineering and Technology
	institute Groningen (ENTEG), University of Groningen, 9747 AG, the Netherlands,
    	{\tt\small n.monshizadeh@rug.nl, c.de.persis@rug.nl}}%
\thanks{This work was supported by the STW perspective program ``Robust Design of Cyber-physical Systems" under the auspices of the project ``Energy Autonomous Smart Microgrids".}
}
\maketitle
\begin{abstract}
%Coping with lossy distribution lines has been an obstacle in control and stability of power networks. Often, imposing an inductive output impedance to the sources has been considered as a means to justify the commonly adopted assumption of purely inductive lines, which in turn facilitates the construction of energy-based Lyapunov functions. 
%Furthermore, additional output impedances are widely used for various purposes.
%, and investigate the extent to which the output impedance can affect the behavior of the distribution lines seen from the voltage source outputs. 
Output impedances are inherent elements of power sources in the electrical grids. In this paper, we give an answer to the following question: What is the effect of output impedances on the inductivity of the power network? To address this question, we propose a measure to evaluate the inductivity of a power grid, and we compute this measure for various types of output impedances. Following this computation, it turns out that network inductivity highly depends on the algebraic connectivity of the network. By exploiting the derived expressions of the proposed measure, one can tune the output impedances in order to enforce a desired level of inductivity on the power system. Furthermore, the results show that the more ``connected" the network is, the more the output impedances diffuse into the network. Finally, using Kron reduction, we provide examples that demonstrate the utility and validity of the method.
\end{abstract}

\renewcommand\IEEEkeywordsname{Index Terms}

\section{Introduction} \label{s:Intro}
%\IEEEPARstart{A}{}microgrid is a small power network consisting of distributed sources and loads which can be seen as one entity from the large area electrical grid. 
\IEEEPARstart{O}{utput} impedance is an important and inevitable
%, and in cases desirable 
element of any power producing device, such as synchronous generators and inverters. Synchronous generators typically possess a highly  inductive output impedance according to their large stator coils, and are prevalently modeled by a voltage source behind an inductance. Similarly, inverters have an inductive output impedance due to the low pass filter in the output, which is necessary to eliminate the high frequencies of the modulation signal.
%In the case of a fault occurring in the main network, such an entity can become isolated and go to the so-called \textit{islanded mode}. In this situation, since few loads and sources are involved, the power consumption and generation becomes highly erratic. To control such a system in terms of voltage control and frequency stabilization, many methods have been proposed during the last few decades. Among these, \textit{droop controllers} have been the main and highly-accepted solution, however there are still on-going researches and debates over this method in order to improve its performance and overcome its limitations \cite{Olivares2014}. In view of the stability analysis and performance of these controllers, coping with lossy lines is perhaps the most challenging problem \cite{droopfail1,droopfail2,droopfail3,droopfail4,droopfail5}.  While energy-based Lyapunov functions can be constructed in the lossless case, such construction is far from trivial when power line losses are taken into account \cite{Schiffer2014, Nima2015, Nima2015_2, Nima2016_1, Nima2017_1, Varaiya1985}. 
%\nmargin{to ask more references from Claudio about lossy Lyapunov based analysis}   

\begin{figure}
\centering
\includegraphics[width=5.5cm]{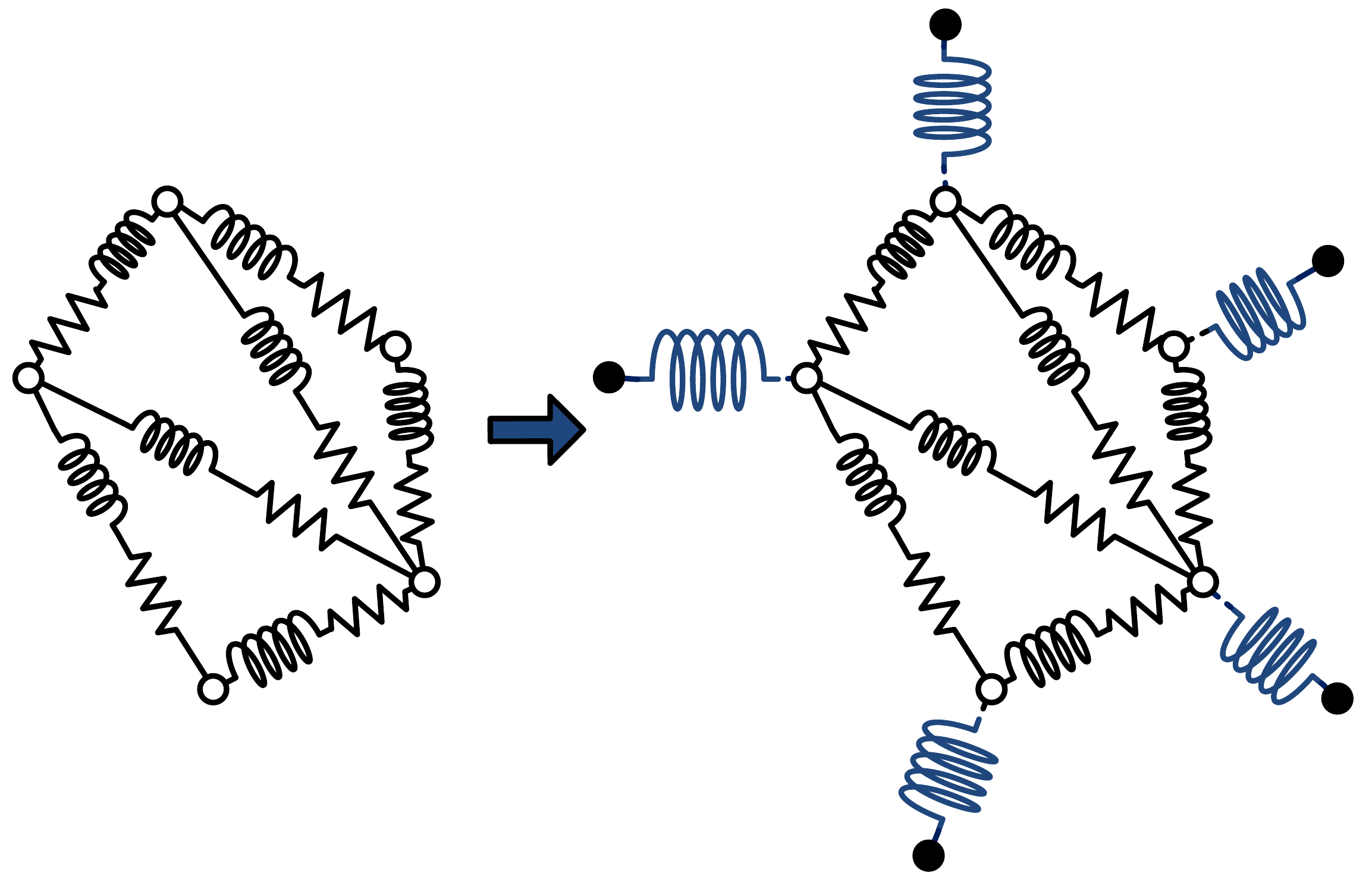}
\caption{Inductive outputs are typically added to the sources in order to assume inductive lines for the resulting network.}
\label{fig:graph}
\vspace{-0.4cm}
\end{figure}

There are motives to add an impedance to the inherent output impedance of the inverters, one of the most important of which is to enhance the performance of droop controllers in a lossy network. Droop controllers show a better performance in a dominantly inductive network (or analogously in dominantly resistive networks for the case of inverse-droop controllers) \cite{TPWRS1,droopfail1,droopfail2,droopfail3,droopfail4,droopfail5,Li2009} (see Figure \ref{fig:graph}). 
The additional output impedance is also employed to improve stability and correct the load sharing error \cite{TPWRS3,droopfail2},\cite{Mahmood2015,Ramin2016}, supply harmonics to nonlinear loads \cite{Brabandere2007},\cite{droopfail1}, \cite{V0}, share current among sources resilient to parameters mismatch and synchronization error \cite{Chiang2001}, decrease sensitivity to line impedance unbalances \cite{TPWRS2,droopfail3},\cite{Guerrero2008}, reduce the circulating currents \cite{Yao2011}, limit output current during voltage sags \cite{Sags}, minimize circulating power \cite{Kim2011}, and damp the $LC$ resonance in the output filter \cite{Li2009}. In most of these methods, to avoid the costs and large size of an additional physical element, a \textit{virtual output impedance} is employed, where the electrical behavior of a desired output impedance is simulated by the inverter controller block. 

Although an inductive output impedance, either resulting from the inherent output filter or the added output impedance, is considered as a means to regulate the inductive behavior of the resulting network, there is a lack of theoretical analysis to verify the feasibility of this method and to quantify the effect of the output impedances on the network inductivity/resistivity. Note that the output impedance cannot be chosen arbitrarily large, since a large impedance substantially boosts the voltage sensitivity to current fluctuations, and results in high frequency noise amplification \cite{Li2009}. Furthermore, there is the fundamental challenge of quantifying inductivity/resistivity of a network, which is nontrivial unless the overall network has uniform line characteristics (homogeneous). This is not the case here as the augmented network will be nonuniform (heterogeneous) even if the initial network is.
%An exception to this, is the network where the distribution lines are assumed to be made of the same material and hence share the same inductivity to resistivity ratio. The lines with such a property are referred to as \textit{homogeneous} lines. In that case, the resulting reduced model can be realized again with homogeneous $RL$ links \cite{Caliskan2012}, \cite{Caliskan2014}. However in our setup, as the output impedances have different inductivity to resistivity ratio compared to the homogeneous lines, the overall network is not homogeneous anymore. Hence, a different and novel treatment is required to evaluate the inductivity behavior of the overall network. 

In this paper, we examine the effect of the output impedances on a homogeneous power distribution grid by proposing a quantitative measure for the inductivity of the resulting heterogeneous network. Similarly, a dual measure is defined for its resistivity. Based on these measures, we show that the network topology plays a major role in the diffusion of the output impedance into the network. Furthermore, we exploit the proposed measures to maximize the effect of the added output impedances on the network inductivity/resistivity. We demonstrate the validity and practicality of the proposed method on various examples and special cases.

The structure of the paper is as follows: In Section \ref{s:method}, the notions of Network Inductivity Ratio ($\Psi_{\rm NIR}$) and Network Resistivity Ratio ($\Psi_{\rm NRR}$) are proposed. In Section \ref{s:Calculating}, the proposed measures are analytically computed for various cases of output inductors and resistors. 
%Section \ref{s:optimization} includes methods for choosing the optimal values of the output impedances are investigated.
%, such that their effect on the network inductivity measure is maximized. 
In Section \ref{s:examples} the proposed measure is evaluated with the Kron reduction in the phasor domain. Finally, Section \ref{s:concl} is devoted to conclusions.
\section{Measure Definition}\label{s:method}
%\n{
%Consider the graph with the added output impedances as depicted in Figure \ref{fig:graph}. As the first step, we reduce the augmented graph to a graph consisting only of the outer nodes. This reduction is through the elimination of the internal nodes (light green nodes in Figure \ref{fig:graph}) and is known as \textit{Kron Reduction} \cite{Kron}. 
%As a result, there can be no graph synthesized with the new reduced graph description with resistors and inductors, 
%for a novel treatment  of the problem in a different way.
%}
%\todoing{to continue with pooya: kron not always possible!}
Consider an electrical network with an arbitrary topology, where we assume that all the sources and loads are connected to the grid via power converter devices (inverters) \cite{Kale2015} {(later in Section \ref{s:Calculating}, we show how to relax this assumption.)}.
 The network of this grid is represented by a connected and weighted undirected graph $\mathcal{G}(\mathcal{V},\mathcal{E}, \Gamma ) $. The nodes $\mathcal{V}= \{ 1,...,n \} $ represent the inverters, and the edge set $\mathcal{E}$  accounts for the distribution lines. The total number of edges is denoted by $m$, i.e., $|\mathcal{E}|=m$. The edge weights are collected in the diagonal matrix $\Gamma$, and will be specified later.
%made precise later., i.e., $\Gamma_{kk}$ denotes the weight of the $k^{th}$ edge of the graph, for each $k $ 
%\n{Furthermore, $\Gamma=\{1,\cdots,\,m\}$ represents the weights. In this notation, edges have the weight $\tau^{-1}_{ij}$, where $\tau_{ij}$ is the distance between nodes $i$ and $j$.}

 For an undirected graph $ \mathcal{G}$, the incidence matrix $B$ is obtained by assigning an arbitrary orientation to the edges of $\mathcal{G}$ and defining
\begin{align*}
b_{ik}=
\begin{cases}
+1 & \text{if } i \text{ is the tail of edge k}
\\
-1 & \text{if } i \text{ is the head of edge k}
\\
0 & \text{otherwise}
\end{cases}
\end{align*}
with $  b_{ik} $ being the $ (i, k)$th element of $ B $. 

We start our analysis with the voltages across the edges of the graph $\mathcal{G}$. We restrict this analysis to the low/medium voltage networks with short line lengths\footnote{A power line is defined as a short-length line if its length is less than 80 km \cite{short1}.}, where the shunt capacitance of the line (pi) model can be neglected \cite[Ch.13]{short1}, \cite[App.1]{short2}, \cite[Ch.6]{short3}. Now let $R_e\in \R^{m\times m}$ and $L_e\in \R^{m\times m}$ be the diagonal matrices with the line resistances and inductances on their diagonal, respectively. We have
%\begin{align*}
%R_{e_i}I_{e_i}+L_{e_i}\dot{I}_{e_i}= V_{e_i}\;,
%\end{align*}
%where $R_{e_i}$ and $L_{e_i}$ are the resistance and inductance of edge $i$. The compact form reads as
\begin{align}\label{e:incidence}
R_eI_e+L_e\dot{I}_e= B^\top V\;,
\end{align}
where $I_e \in \R^m$ denotes the current flowing through the edges. The orientation of the currents is taken in agreement with that of the incidence matrix. 
 The vector $V\in \R^n$ indicates the voltages at the nodes. 
Let $\tau_k$ denote the physical distance between nodes $i$ and $j$, for each edge $k \sim \{i, j\}$. We assume that the network is homogeneous, i.e. the distribution lines are made of the same material and possess the same resistance and inductance per length: $$r=\frac{R_{e_k}}{\tau_{e_k}},\;\;\; l=\frac{L_{e_k}}{\tau_{e_k}},\hspace{1cm} k=\{1,\cdots,m\}\;.$$
Now, let the weight matrix $\Gamma$ be specified as
\begin{equation}\label{e:Gamma}
\Gamma={\rm diag}(\gamma):={\rm diag}(\tau_1^{-1}, \tau_2^{-1}, \cdots,\tau_m^{-1})\;.
\end{equation}
We can rewrite \eqref{e:incidence} as \cite{Caliskan2014}
\begin{align*}
rI_e+\ell\dot{I}_e=\Gamma B^\top V.
\end{align*}
%where $\Gamma={\rm diag}(\tau_1^{-1},\;\cdots,\tau_m^{-1})$ with $\tau_i$ denoting the length of link $i$. 
Hence,
%\begin{align*}
$rBI_e+\ell B\dot{I}_e=B\Gamma B^\top V,$
%\end{align*}
and
\begin{align}\label{e:Laplacian}
r I +\ell\dot{I}=\la V\;,
\end{align}
where $I:=BI_e$ is the vector of nodal current injections. The matrix $\la=B\Gamma B^\top $ is the Laplacian matrix of the graph $\mathcal{G}(\mathcal{V},\mathcal{E}, \Gamma )$ with the weight matrix $\Gamma$.
% (independent of the orientation). 

Note that, as the network \eqref{e:Laplacian} is homogeneous, its inductivity behavior is simply determined by the ratio $\frac{\ell}{r}$. 
However, clearly, network homogeneity will be lost once the output impedances are augmented to the network. 
This makes the problem of determining network inductivity nontrivial and challenging.
To cope with the heterogeneity resulting from the  addition of the output impedances, we need to depart from the homogeneous form \eqref{e:Laplacian}, and develop new means to assess the network inductivity. To this end, we consider the more general representation
%, the one of \eqref{e:define} cannot be trivially quantified. 
%simply determined by the ratio $\frac{\ell}{r}$. 
%, the one of \eqref{e:define} cannot be trivially quantified. 
%
%
%The representation \eqref{e:Laplacian} will be used afterwards once the output impedances are augmented to the network. 
%Then, a reasonable approach to quantify the influence of the output impedance on the inductivity of the overall network, is to look at the edges of the resulting graph. However, as 
%It will be shown later in Section \ref{s:Calculating}, except some special cases, the overall model cannot be described by \eqref{e:Laplacian} (see e.g. \eqref{e:rlmodel}), and hence there can be no RL network built based on the resulting network description. 
% However, this requires that an electrical network can be built such that its mathematical description matches that of the reduced model.
%While this is always possible in purely resistive networks \cite{Arjan2010}, such synthesis is generally not feasible in an $RL$ network \cite{Willems2010}. 
%Consequently, the overall model can not be described by \eqref{e:Laplacian}. 
%To tackle this problem, we depart from the equation \eqref{e:Laplacian} and consider the more general case in which the network is described by 
\begin{align}\label{e:define}
RI+L\dot{I}=\la V_o\;,
\end{align}
where $V_o\in \R^n$ is the vector of voltages of the augmented nodes ({black} nodes in Figure \ref{fig:graph}), and $R\in \R^{n \times n}$ and $L\in \R^{n \times n}$ are matrices associated closely with the resistances and inductances of the lines, respectively. We will show that the overall network after the addition of the output impedances, can be described by \eqref{e:define}. Note that this description cannot necessarily be realized with passive RL elements. 
%Hence we try to analyze the inductivity ratio through a different property.
Therefore, while the inductivity behavior of the homogeneous network \eqref{e:Laplacian} is simply determined by the ratio $\frac{\ell}{r}$, the one of \eqref{e:define} cannot be trivially quantified. 

The idea here is to promote the rate of convergence as a suitable metric quantifying the inductivity/resistivity of the network. For the network dynamics in \eqref{e:Laplacian}, the rate of convergence of the solutions is determined by the ratio $\frac{r}{\ell}$. The more inductive the lines are, the slower the rate of convergence is. Now, we seek for a similar property in \eqref{e:define}.  Notice that the solutions of (3) are damped with corresponding eigenvalues of $L^{-1}R$. 
Throughout the paper, we assume the following property:
\begin{assumption}\label{a:stable}
The eigenvalues of the matrix $L^{-1}R$ are all positive and real. 
 \end{assumption}
 It will be shown that Assumption \ref{a:stable} is satisfied for all the cases 
 considered in this paper. 
%\begin{remark}
%	Any heterogeneous electrical network with a tree graph can be modeled by \eqref{e:define}, as from \eqref{e:incidence} we have
%	$$
%	R_e(B^\top B)^{-1}B^\top (BI_e)+L_e(B^\top B)^{-1}B^\top (B\dot I_e)=B^\top  V\;.
%	$$	
%	Hence
%	$$
%	R_\T I+L_\T\dot{I}=\la V\;,
%	$$
%	where $R_\T=BR_e(B^\top B)^{-1}B^\top ,\;L_\T=BL_e(B^\top B)^{-1}B^\top $. Therefore, the proposed measures can also be used to quantify the inductivity of a heterogeneous network with a tree topology.
%\end{remark}

\begin{figure}
	\centering
	\includegraphics[width=8.8cm]{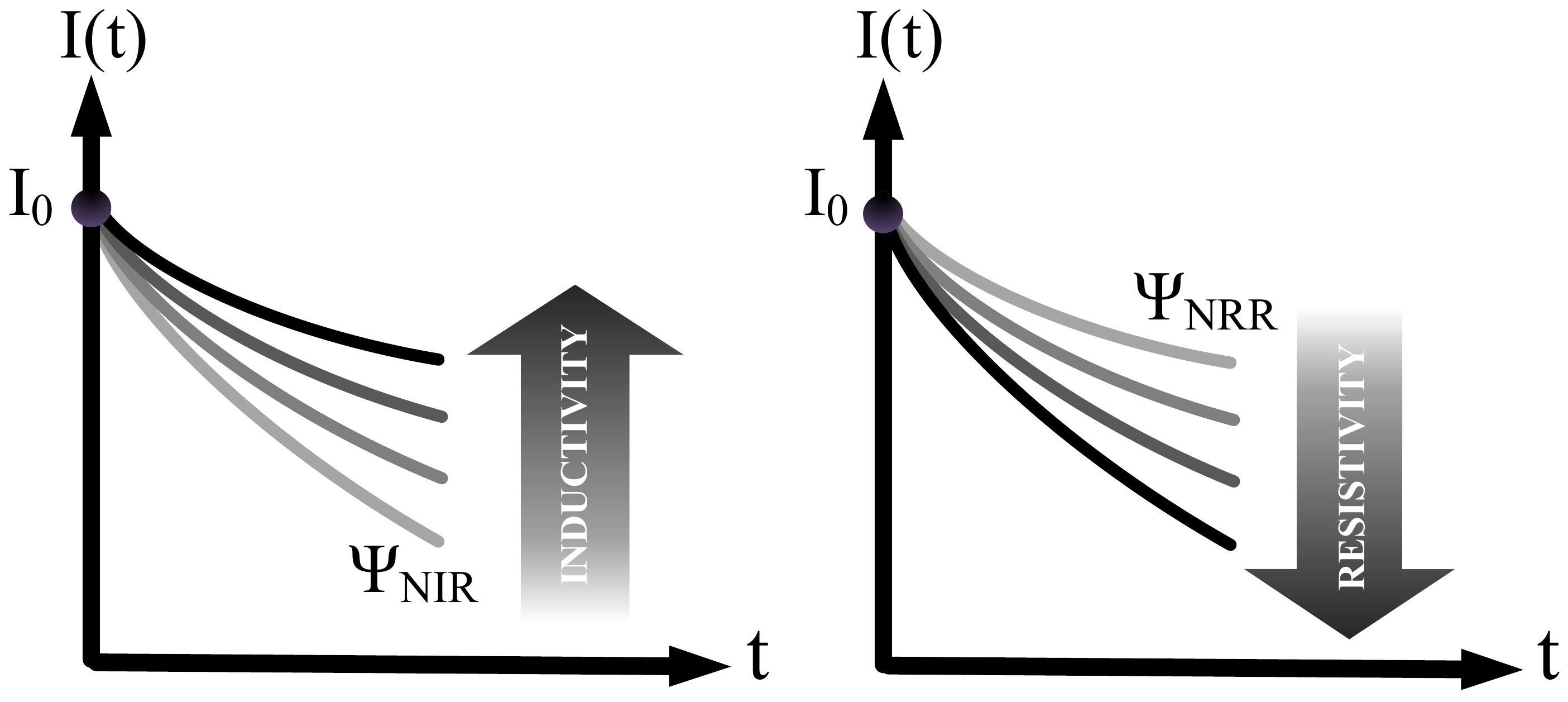}
	\caption{Worst cases are selected for inductivity and resistivity measures.}
	\label{fig:current}
\end{figure}

Figure \ref{fig:current} sketches the behavior of homogeneous solutions of \eqref{e:define}. 
Among all the solutions, we choose the fastest one as our measure for inductivity, and the slowest one for resistivity of the network.  
%; i.e. the most damped solution as the inductivity measure, and the least damped solution as the resistivity ratio. 
Opting for these worst case scenarios allows us to guarantee a prescribed inductivity or resistivity ratio by proper design of output impedances. These choices are formalized in the following definitions.
\begin{definition}\label{d:NIR}
	Let $I(t,I_0)$ denote the 
%	unique solution of \eqref{e:define} with uniform node voltages $V=\1v^*$
	homogeneous solution of \eqref{e:define} for an initial condition $I_0\in \im B$. Let the set $M_L\subseteq \R^+$ be 
	given by 
	\begin{align*}
	\begin{aligned}
	M_L&:= \{\sigma \in \R^+\;|\;\; \exists \mu \;\text{s.t.}\\
&
	\| I(t, I_0)\| \geq \mu e^{-\sigma t}\| I_{0}\|,
\; \forall t\in \R^+, \; \forall I_{0} \in \im B \}.
	\end{aligned}
	\end{align*}Then we define the \textit{Network Inductivity Ratio} (NIR) as
	%\begin{align}\label{e:NIR}
	\[\Psi_{\rm NIR}:=\frac{1}{\inf(M_L)}.\]
	%\end{align}
	{\hspace{\fill} $\square$\noindent}
%{\hspace*{\fill} $\square$\noindent}
\end{definition}

\begin{definition}
Let $I(t,I_0)$ denote the 
%	unique solution of \eqref{e:define} with uniform node voltages $V=\1v^*$
homogeneous solution of \eqref{e:define} for an initial condition $I_0\in \im B$. Let the set $M_R\subseteq \R^+$ be given by
\begin{align*}
\begin{aligned}
M_R&:= \{\sigma \in \R^+\;|\;\; \exists \mu \;\text{s.t.} \\
&\| I(t, I_0)\| \leq \mu e^{-\sigma t}\| I_{0}\|,\;\forall t\in \R^+, \; \forall I_{0} \in \im B \}.
\end{aligned}
\end{align*}
We define the \textit{Network Resistivity Ratio} (NRR) as
%\begin{align}%\label{e:NRR}
\[\Psi_{\rm NRR}:=\sup(M_R).
\]
%\end{align}
{\hspace*{\fill} $\square$\noindent}
\end{definition}
Note that the set $M_L$ is bounded from below and $M_R$ is bounded from above by definition and Assumption \ref{a:stable}. 
%Moreover, these sets are non-empty and well-defined since $R$ and $L$ are positive definite matrices.
Interestingly, in case of the homogeneous network \eqref{e:Laplacian}, i.e. without output impedances, we have $\Psi_{\rm NIR}=\frac{\ell}{r}$ and $\Psi_{\rm NRR}=\frac{r}{\ell}$, which are natural measures to reflect the inductivity and resistivity of an $RL$ homogeneous network.% \eqref{e:Laplacian}.
\section{Calculating the Network Inductivity/Resistivity Measure ($\Psi_{\rm NIR}$/$\Psi_{\rm NRR}$)}\label{s:Calculating}
In this section, based on Definitions 1 and 2, we compute the network inductivity/resistivity ratio for both cases of uniform and nonuniform output impedances.
\subsection{Uniform Output Impedances}
In most cases of practical interest, the output impedance consists of both inductive and resistive elements. We investigate the effect of the addition of such output impedances on the network inductivity ratio. 
The change in network resistivity ratio can be studied similarly, and thus is omitted here. 
%scenario where we aim at making the network dominantly inductive by adding . 
%The analysis for \n{the case of} \sout{aiming at} a dominantly resistive network is analogous.
Consider the uniform output impedances with the inductive part $\ell_o$ and the resistive component $r_o$ (in series), added to the network \eqref{e:Laplacian}. Note that the injected currents $I$ now pass through the output impedances, as shown in Figure \ref{fig:3D}. Clearly,
we have
\begin{align}\label{e:1}
V=V_o-r_oI-\ell_o\dot{I} \;.
\end{align}
%Recall that the injected currents 
%satisfy the equation \eqref{e:Laplacian}. 
Having \eqref{e:Laplacian} and \eqref{e:1}, the overall network can be described as
\begin{align}\label{e:rlmodel}
(r_o\la +r\I)I+(\ell_o\la+\ell\I)\dot{I}=\la  V_o\;,
\end{align}
where $\I \in \R^{n\times n}$ denotes the identity matrix, and $\mathcal{L}$ is the Laplacian matrix of $\mathcal{G}$ as before.
In view of equation \eqref{e:define}, the matrices $R$ and $L$ are given by $R=r_o\la +r\I$ and $L=\ell_o\la+\ell\I$, respectively. As both matrices are positive definite, the eigenvalues of the product $L^{-1}R$ are all positive and real, see \cite[Ch. 7]{Horn2012}. Hence, Assumption \ref{a:stable} is satisfied. To calculate the measure $\Psi_{\rm NIR}$ for the inductivity of the resulting network, we investigate the convergence rates of the homogeneous solution of \eqref{e:rlmodel}. This brings us to the following theorem:
\begin{figure}
	\centering
	\includegraphics[trim={2cm 0 2cm 0},clip,width=5.0cm]{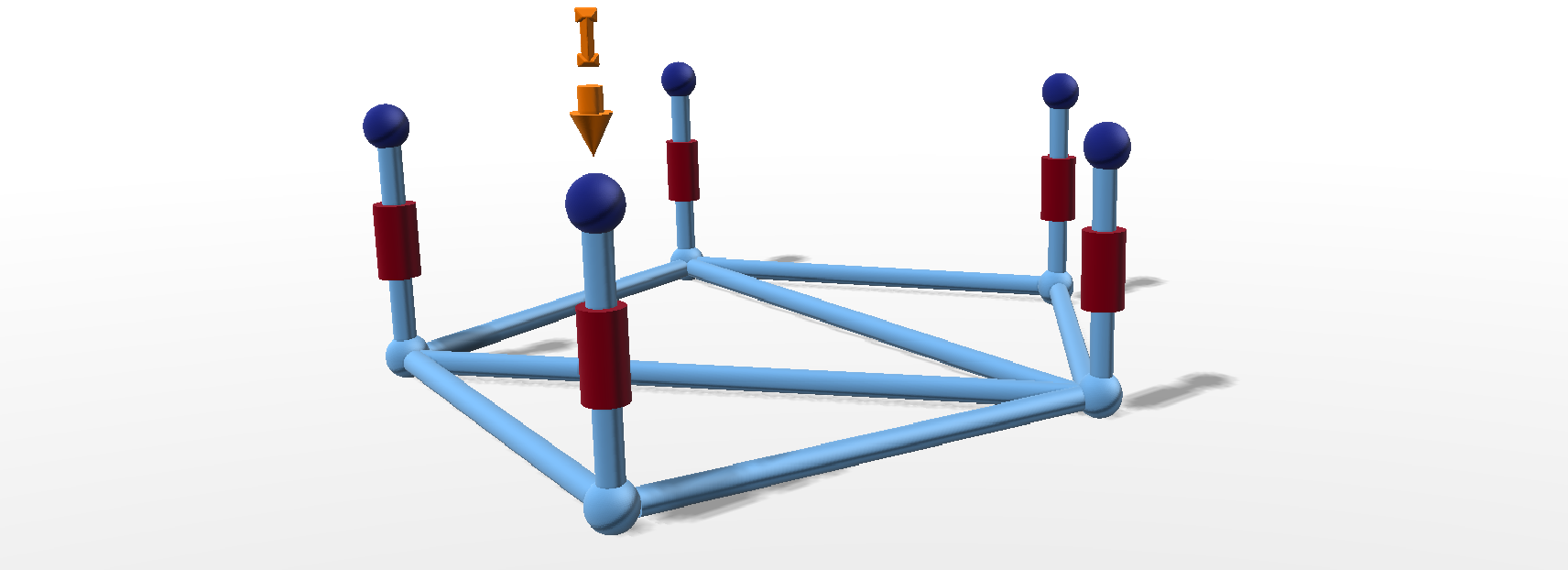}
	\caption{The injected currents at the nodes of the original graph pass through the added output impedance.}
	\label{fig:3D}
\end{figure}
\begin{theorem}\label{t:rl}
	Consider a homogeneous network \eqref{e:Laplacian} with the resistance per length unit $r$ and inductance per length unit $\ell$. Suppose that an output resistance $r_o$ and an output inductance $\ell_o$ are attached in series to each node. Assume that $\frac{r_o}{\ell_o}<\frac{r}{\ell}$. Then the network inductivity ratio is given by
	\begin{equation}\label{e:nima-NIR}
	\Psi_{\rm NIR}=\frac{\ell_o\lambda_2+\ell}{r_o\lambda_2+r}\;,
	\end{equation}
	where $\lambda_2$ is the \textit{algebraic connectivity}\footnote{The algebraic connectivity of either a directed or an undirected graph $\mathcal{G}$ is defined as the second smallest eigenvalue of the Laplacian matrix throughout the paper. Note that the smallest eigenvalue is $0$.} of the graph $\mathcal{G}(\mathcal{V},\mathcal{E}, \Gamma )$.
\end{theorem}
\begin{IEEEproof}
	The homogeneous solution is
	\begin{align*}
	I(t)&=e^{-(r_o\la+r\I)(\ell_o\la+\ell\I)^{-1}t}I_{0}\;.
	\end{align*}
	The Laplacian matrix can be decomposed as $\la=\U^\top \Lambda\U$. Here, $U$ is the matrix of eigenvectors and $\Lambda=\diag\{\lambda_1,\lambda_2,\cdots,\lambda_n\}$ where $\lambda_1<\lambda_2<\cdots<\lambda_n$ are the eigenvalues of the matrix $\la$. Note that $\lambda_1=0$. We have
	\begin{align*}
	I(t)&=e^{-\U(r_o\Lambda+r\I)\U^\top \big(\U(\ell_o\Lambda+\ell\I)\U^\top \big)^{-1}t}I_{0}
	\\
	&=\U e^{-(r_o\Lambda+r\I)(\ell_o\Lambda+\ell\I)^{-1}t}\U^\top  I_{0}
	\\
	&=\begin{bmatrix}
	\frac{1}{\sqrt{n}}\1 \;\;\; \tilde{\U}
	\end{bmatrix} e^{-
		\begin{bmatrix}
		\frac{r}{\ell} && 0
		\\
		0_{(n-1)\times 1} && \tilde{\Lambda}
		\end{bmatrix}t}
	\begin{bmatrix}
	\frac{1}{\sqrt{n}}\1^\top 
	\\
	\tilde{\U}^\top 
	\end{bmatrix}
	I_{0}
	\;,
	\end{align*}
	where $\tilde{\Lambda}={\rm diag}\{\frac{r_o\lambda_2+r}{\ell_o\lambda_2+\ell},\cdots,\frac{r_o\lambda_n+r}{\ell_o\lambda_n+\ell}\}$. 
	Noting that $\U$ is unitary and by the Kirchhoff Law, $\1^\top I_{0}=0$, we have
	\begin{align*}
	&I(t)=\tilde{\U} e^{-\tilde{\Lambda}t}\;\tilde{\U}^\top  I_{0}
	=(\sum_{i=1}^{n-1}e^{-\tilde{\lambda}_it}\;\tilde{\U}_i \tilde{\U}_i ^\top  ) (\sum_{i=1}^{n-1}\alpha_i \tilde{\U}_i)
	\\
	&\qquad =\sum_{i=1}^{n-1}\alpha_i e^{-\tilde{\lambda}_it}\;\tilde{\U}_i \;,
	\end{align*}
	where $\tilde{\U}_i$ denotes the $i$th column of $\tilde{\U}$, and we used again $\1^\top I_{0}=0$ to write $I_{0}$ as the linear combination $$I_0=\sum_{i=1}^{n-1}\alpha_i \tilde{\U}_i\;.$$
	%, and $\tilde{\lambda}_i=\ell_o\lambda_{i+1}+\ell$
	Hence
	\begin{align}\label{e:I(t)}
	\lVert I(t) \lVert^2=\sum_{i=1}^{n-1}\alpha_i^2 e^{-2\tilde{\lambda}_it} \;.
	\end{align}
	Having  $\frac{r_o}{\ell_o}<\frac{r}{\ell}$, it is straightforward to see that $$\frac{r_o\lambda_2+r}{\ell_o\lambda_2+\ell}\geq\frac{r_o\lambda_i+r}{\ell_o\lambda_i+\ell},\;\;\;\forall i\;\;.$$
	and bearing in mind that $\lVert I_{0} \lVert^2=\sum_{i=1}^{n-1}\alpha_i^2 $, we conclude that
	\begin{align}\label{e:geq}
	&\lVert I(t) \lVert \geq e^{-\frac{r_o\lambda_2+r}{\ell_o\lambda_2+\ell}t}\lVert I_{0} \lVert \;,
	\end{align}
	which yields $\Psi_{\rm NIR}=\frac{\ell_o\lambda_2+\ell}{r_o\lambda_2+r}$.
	Note that \eqref{e:geq} holds with equality in case $I_{0}$ belongs to the span of the corresponding eigenvector of the second smallest eigenvalue of the Laplacian matrix $\la$. This completes the proof.
\end{IEEEproof}

Theorem \ref{t:rl} provides a compact and easily computable expression which quantifies the network inductivity behavior. 
Moreover,  the expression \eqref{e:nima-NIR} is an easy-to-use measure that can be exploited to choose the output impedances in order to impose a desired degree of inductivity on the network. 
The only information required is the line parameters $r$, and $\ell$, and the algebraic connectivity of the network.

	Algebraic connectivity is a measure of connectivity of the weighted graph $\mathcal{G}$, which depends on both the density of the edges and the weights (inverse of the lines lengths). Hence, Theorem \ref{t:rl} reveals the fact that: \textit{``The more connected the network is, the more the output impedance diffuses into the network."}.
	
		The algebraic connectivity of the network can be estimated through distributed methods \cite{Distributed1}, \cite{Distributed2}. Furthermore, line parameters (resistance and inductance) can be identified through PMUs (Phase Measurement Units) \cite{PMU1} \cite{PMU2} \cite{PMU3}. Therefore, our proposed measure can be calculated in a distributed manner. 
\begin{remark}
	In case the resistance part of the output impedance is negligible, i.e $r_o=0$,  the network inductivity ratio reduces to
	\[\Psi_{\rm NIR}=\frac{\ell_o\lambda_2+\ell}{r}\;,\]
	
	In case $\frac{r_o}{\ell_o}>\frac{r}{\ell}$, the network inductivity ratio will be given by
	\[
	\Psi_{\rm NIR}=\frac{\ell_o\lambda_{\max}+\ell}{r_o\lambda_{\max}+r}\;,
	\]
	where $\lambda_{\max}$ is the largest eigenvalue of the Laplacian matrix of $\mathcal{G}$. Furthermore, if $\frac{r_o}{\ell_o}=\frac{r}{\ell}$, then $\tilde{\Lambda}=\frac{r}{\ell}\I$ and $\Psi_{\rm NIR}=\frac{r}{\ell}$.
	However, the condition $\frac{r_o}{\ell_o}<\frac{r}{\ell}$ assumed in Theorem \ref{t:rl} is more relevant since the resistance $r_o$ of the inductive output impedance is typically small.
	\ER
\end{remark}
%\begin{IEEEproof}
%	This is a special case of Theorem \ref{t:rl} with $r_o=0$.
%\end{IEEEproof}
%\subsection{Resistive Output Impedance}
%Consider the output resistors $r_o$ added to the network described by \eqref{e:Laplacian}. The overall network can be described as
%\begin{align}\label{e:rmodel}
%(r_o\la +r\I)I+\ell\dot{I}=\la  V_o\;.
%\end{align}
%Compared to the equation \eqref{e:define}, here $R=r_o\la +r\I$ and $L=\ell\I$. Hence, similar to the inductive case, the matrix $L^{-1}R$ has positive real eigenvalues, and the network resistivity ratio in this case is obtained as follows:

As mentioned in Section \ref{s:Intro}, in low-voltage microgrids where the lines are dominantly resistive, the inverse-droop method is employed. In this case, a purely resistive output impedance is of advantage \cite{Resistive}.
\begin{corollary}
Consider a homogeneous distribution network with the resistance per length unit $r$, inductance per length unit $\ell$, and output inductors $\ell_o$. Then the network resistivity ratio is given by
\[
\Psi_{\rm NRR}=\frac{r_o\lambda_2+r}{\ell}\;,
\]
where $\lambda_2$ is the \textit{algebraic connectivity} of the graph $\mathcal{G}(\mathcal{V},\mathcal{E}, \Gamma )$.
\end{corollary}
\begin{IEEEproof}
The proof can be constructed in an analogous way to the proof of Theorem \ref{t:rl} and is therefore omitted.
\end{IEEEproof}
	\begin{remark}
		The homogeneity assumption is ubiquitous in the literature  of power network analysis (see e.g. \cite{TPWRS4, Caliskan2012,Caliskan2014,Zhong2012,Munz2014,Sinha2016,Nima2018}). 
Here we show briefly how the results can be extended to the case of a heterogeneous network. 
\\
Using \eqref{e:incidence} and \eqref{e:1} together with $I=BI_e$ we have
\begin{align}\label{e:het}
&R_eI_e+L_e\dot{I}_e= B^\top (V_o-r_oI-\ell_o\dot{I})
\\
&(r_o\la_e+R_e)I_e+(\ell_o\la_e+L_e)\dot I_e= B^\top V_o
\;,
\end{align}
		where $\la_e:=B^\top B$ is the \textit{edge Laplacian}. Note that since the edge Laplacian is symmetric and positive semi-definite, the matrix $(\ell_o\la_e+L_e)^{-1}(r_o\la_e+R_e)$ has positive real eigenvalues, and hence analogously to the Definition \ref{d:NIR} and Theorem \ref{t:rl}, the network inductivity ratio can be defined and computed for the case of a heterogeneous network with additional output impedances.	However, computing closed-form expressions for the proposed inductivity metric will become more challenging, and is not pursued in this work.
\ER

\end{remark}
\begin{remark}
	One of the main desired features in microgrids is plug-and-play capability for planning and connection of the new sources. In most cases, a new node connects to the network initially through few edges. This results in a decrease in the algebraic connectivity of the overall network, e.g., as shown in \cite{Grone1990},
	adding a pendant vertex and edge to a graph does not increase the
	algebraic connectivity. Therefore, for plug-and-play capability, larger output impedances should be employed in the network to compensate the possible drop in the algebraic connectivity, and thus the network inductivity ratio, resulting from attaching new nodes to the network. In some special cases, such as uniform line lengths, the additional required output impedances can be estimated using lower bounds on the algebraic connectivity; see \cite{Mohar1992} and \cite{Abreu2007} for more details on algebraic connectivity and its lower and upper bounds in various graphs.
	\ER
\end{remark}
\subsubsection{Case Study} \textit{Identical Line Lengths}\\
\vspace{-3mm}

Recall that the notion of network inductivity ratio allows us to quantify the inductivity behavior of the network, while 
the model \eqref{e:rlmodel}, in general, cannot be synthesized with $RL$ elements only. 
A notable special case where the model \eqref{e:rlmodel} can be realized with $RL$ elements is a complete graph with identical line lengths. Although such case is improbable in practice, it provides an example to assess the validity and credibility of the introduced measures. Interestingly, $\Psi_{\rm NIR}$ matches precisely the inductance to resistance ratio of the lines of the synthesized network in this case:
\begin{theorem}\label{t:complete graph}
	Consider a network with a uniform complete graph where all the edges have the length $\tau$. Suppose that the lines have inductance $\ell_e\in \R$ and resistance $r_e\in \R$. Attach an output inductance $\ell_o$ in series with a resistance $r_o$ to each node. Then the model of the augmented graph can be equivalently synthesized by a new $RL$ network with identical lines, each with inductance $\ell_c:=n\ell_o+\ell_e$ and resistance $r_c:=nr_o+r_e$, where $n$ denotes the number of nodes. Furthermore, the resulting network inductivity ratio $\Psi_{\rm NIR}$ is equal to  $\frac{\ell_c}{r_c}$.
\end{theorem}
\begin{IEEEproof}
	The nodal injected currents satisfy
	%\begin{align*}
	$rI+\ell\dot{I}=\la V.$
	%\end{align*}
	In this network, $r=\frac{r_e}{\tau}$, $\ell=\frac{\ell_e}{\tau}$, and $\la=\frac{n}{\tau}\Pi$ where $\Pi:=\I-\frac{1}{n}\1\1^\top $. Hence, 
	\begin{align}\label{e:complete}
	r_eI+\ell_e\dot{I}=n\Pi V\;.
	\end{align}
	By appending the output impedance we have
	$V=V_o-r_oI-\ell_o\dot{I}$.
    Hence \eqref{e:complete} modifies to
	\begin{align*}
	(nr_o\Pi+r_e\I)I+(n\ell_o\Pi+\ell_e\I)\dot{I}=n\Pi V_o\;,
	\end{align*}
	which results in
	\begin{align*}
	(n\ell_o\Pi+\ell_e\I)^{-1}(nr_o\Pi+r_e\I)I+\dot{I}=n(n\ell_o\Pi+\ell_e\I)^{-1}\Pi V_o\;.
	\end{align*}
	Since $(n\ell_o\Pi+\ell_e\I)^{-1}=\frac{1}{\ell_e+n\ell_o}\I+\frac{\ell_o}{\ell_e(\ell_e+n\ell_o)}\1\1^\top $, we obtain
	\begin{align}
	(nr_o\Pi+r_e\I)I+(n\ell_o+\ell_e)\dot{I}=n\Pi V_o\;,
	\end{align}
	where we used $\1^\top I=0$ and $\1^\top \Pi=0$. Similarly we have
	\begin{align*}
	I+\ell_c(nr_o\Pi+r_e\I)^{-1}\dot{I}=n(nr_o\Pi+r_e\I)^{-1}\Pi V_o\;,
	\end{align*}
	and hence
%	since $(nr_o\Pi+r_e\I)^{-1}=\frac{1}{r_e+nr_o}\I+\frac{r_o}{r_e(r_e+nr_o)}\1\1^\top $, we get
	%\begin{align}\label{e:complete2}
%	(nr_o+r_e)I+(n\ell_o+\ell_e)\dot{I}=n\Pi V_o\;.
	$r_cI+\ell_c\dot{I}=n\Pi V_o$.
	%\end{align}
	This equation is analogous to \eqref{e:complete} and corresponds to a uniform complete graph with identical line resistance $r_c=nr_o+r_e$ and inductance $\ell_c=n\ell_o+\ell_e$.

Note that the algebraic connectivity of  the weighted Laplacian $\la$ is $\frac{n}{\tau}$. By Theorem \ref{t:rl}, the inductivity ratio is then computed as  $$\Psi_{\rm NIR}=\frac{\frac{n}{\tau}\ell_o+\ell}{\frac{n}{\tau}r_o+r}=\frac{\ell_c}{r_c}\;.$$
\end{IEEEproof}
%\vspace*{-0.5cm}
\subsubsection{Case Study}\textit{Constant Current Loads}\\
\vspace{-3mm}

So far, we have considered loads which are connected via power converters. The same definitions and results can be extended to the case of loads modeled with constant current sinks. Consider the graph $\mathcal{G}(\mathcal{V},\mathcal{E}, \Gamma )$ divided into source ($S$) and load nodes ($L$), and decompose the Laplacian matrix accordingly as
\[
\la=\begin{bmatrix}
\la_{SS} && \la_{SL}
\\
\la_{LS} && \la_{LL}
\end{bmatrix}.
\]
We have
\begin{align}
r I_{S}+ \ell \dot{I}_{S} &= \la_{SS} V_{S}+\la_{SL}V_L \label{e:cc1}
\\
r I_{L}+ \ell \dot{I}_{L} &= \la_{LS} V_{S}+\la_{LL}V_L\;. \label{e:cc2}
\end{align}
Suppose that the load nodes are attached to constant current loads $I_L=-I_L^*$. Then from \eqref{e:cc2} we obtain
$$-r I_{L}^* = \la_{LS} V_{S}+\la_{LL}V_L\;,$$ and therefore 
\begin{align}\label{e:cc3}
-r \la_{LL}^{-1} I_{L}^* -\la_{LL}^{-1}\la_{LS} V_{S}= V_L\;.
\end{align}
Substituting \eqref{e:cc3} into \eqref{e:cc1} yields
$$r I_{S}+ \ell \dot{I}_{S} = \la_{\redu} V_{S}-r\la_{SL} \la_{LL}^{-1} I_{L}^*\;.$$
Here the Scur complement $\la_{\redu}=\la_{SS} -\la_{SL}\la_{LL}^{-1}\la_{LS}$ is again a Laplacian matrix known as the \textit{Kron-reduced} Laplacian \cite{Kron}, \cite{Arjan2010}. Bearing in mind that $V_G=V_o-\ell_o\dot{I}_S-r_o I_S$, the system becomes
\begin{align}\label{e:ccred}
\begin{aligned}
(r \I +r_o \la_{\redu}) I_{S}+ (\ell \I +&\ell_o \la_{\redu}) \dot{I}_{S} 
\\&= \la_{\redu} V_o-r\la_{SL} \la_{LL}^{-1} I_{L}^*\;,
\end{aligned}
\end{align}
and one can repeat the same analysis as above working with $\la_{\redu}$ instead of $\la$. Note that \eqref{e:ccred}  matches the model \eqref{e:define} with the difference of a constant. As this constant term does not affect the homogeneous solution, the network inductivity and resistivity ratios are obtained analogously as before, where the algebraic connectivity is computed based on the Kron reduced Laplacian. 
%Here, since the Laplacian is loopy, smallest eigenvalue should be used instead of the algebraic connectivity.
\ER
\subsection{Non-uniform Output Impedances}
In this section we investigate the case where output inductances with different magnitudes are connected to the network, and we quantify the network inductivity ratio $\Psi_{\rm NIR}$ under this non-uniform addition. The case with non-uniform resistances can be treated in an analogous manner.

For the sake of simplicity, throughout this  subsection, we consider the case where the resistive parts of the output impedances are negligible (see Remark \ref{r:reviewer} for relaxing this assumption). Let $D= \diag (\ell_{o_1},\ell_{o_2},\cdots,\ell_{o_n})$, where $\ell_{o_i}$ is the (nonzero) output inductance connected to the node $i$. We have
\begin{align*}
%\begin{cases}
rI+\ell\dot{I}= \la V, \qquad
V=V_o-D\dot{I},
%\end{cases}
\end{align*}
and hence
\begin{align}\label{e:various}
rI+(\ell\I+\la D) \dot{I}= \la V_o\;.
\end{align}
Note that $\la D$ is similar to $D^{\half} \la D^{\half}$ and therefore has nonnegative real eigenvalues. In view of equation \eqref{e:define}, here $R=r\I$ and $L=\ell\I+\la D$. Hence, the matrix $L^{-1}R$ possesses positive real eigenvalues, and Assumption \ref{a:stable} holds.

\begin{figure}
	\centering
	\includegraphics[width=8cm]{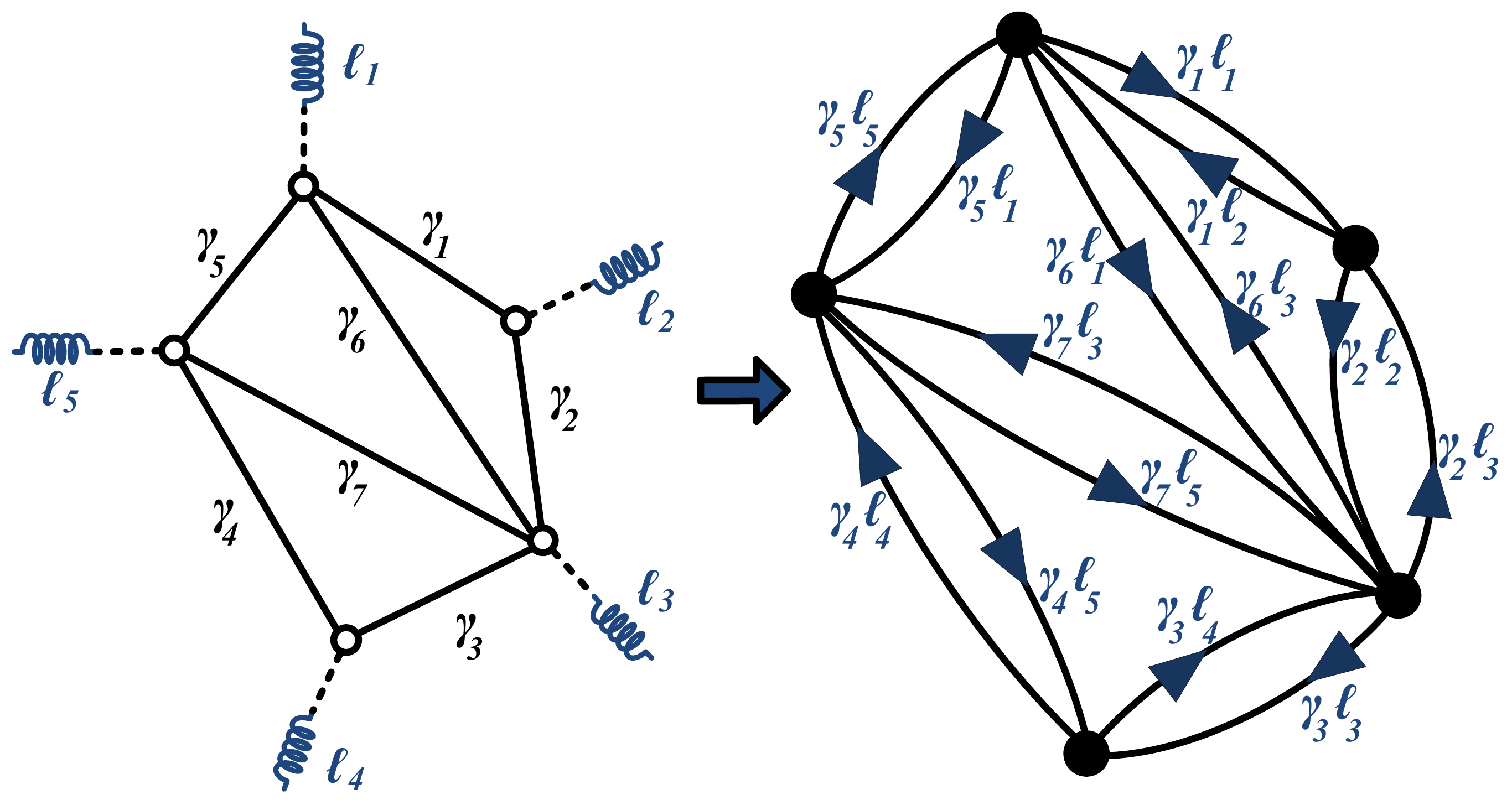}
	\caption{Output inductances appear as weights in the corresponding directed graph, with the Laplacian $D \la$.}
	\label{fig:Lprime}
\end{figure}

The matrix $\la D$ is also similar to $D \la$, which can be interpreted as the (asymmetric) Laplacian matrix of a directed connected graph noted by   $\hat{\mathcal{G}}(\mathcal{V},\hat{\mathcal{E}},\hat \Gamma ) $ with the same nodes as the original graph $\mathcal{V}= \{ 1,...,n \}$, but with directed edges $\hat{\mathcal{E}} \subset \mathcal{V} \times \mathcal{V}$. As shown in Figure \ref{fig:Lprime}, in this representation, for any $(i,j)\in {\mathcal{\hat E}}$, there exists a directed edge from node $i$ to node $j$ with the weight $\ell_{o_i} \tau^{-1}_{ij}$ (recall that $\tau^{-1}_{ij}$ is the weight of the edge $\{i,j\}\in \mathcal{E}$ of the original graph $\mathcal{G}$). Hence, the weight matrix $\hat\Gamma\in \R^{2m\times 2m}$ is the diagonal matrix with the weights $\ell_{o_i} \tau^{-1}_{ij}$ on its diagonal. %given by $\hat\Gamma=D\Gamma$.
Note that the edge set $\mathcal{\hat E}$ is symmetric in the sense that $(i, j)\in \mathcal{\hat E} \Leftrightarrow (j, i)\in \mathcal{\hat E}$, and its cardinality is equal to $2m$. We take advantage of this graph to obtain the network inductivity ratio $\Psi_{\rm NIR}$, as formalized in the following theorem.
%\todoing{If you are using different graphs with different weights, it might be better to work with the explicit notation $\mathcal{G}(\mathcal{V}, \mathcal{E}, {\Gamma})$ from the beginning. (but if this requires major rewriting, we can skip it)}
\begin{theorem}\label{t:LD}
Consider a homogeneous network with the resistance per length unit $r$, inductance per length unit $\ell$, edge lengths $\tau_1,\cdots,\tau_n$, and output inductors $\ell_{o_1},\ell_{o_2},\cdots,\ell_{o_n}$. Then the network inductivity ratio is given by 
\[\Psi_{\rm NIR}=\frac{\lambda_2+\ell}{r}\;,\]
where $\lambda_2$ is the \textit{algebraic connectivity} of the graph $\hat{\mathcal{G}}(\mathcal{V},\hat{\mathcal{E}} ,\hat \Gamma)$ defined above.
%with the weight of the oriented edge $(i,j)$ given by $\ell_{o_i}\tau^{-1}_{ij}$.
\end{theorem}
%\todoing{refer to the graph}
\begin{IEEEproof}
Let $\la'=D^{\half}\la D^{\half}$. The homogeneous solution to \eqref{e:various} is 
\begin{align*}
I(t)&=e^{-r(\ell\I+\la D)^{-1}t}I_0
\\&=D^{-\half}e^{-rD^{\half}(\ell \I+\la D)^{-1}D^{-\half}t}D^{\half}I_0
\\
&=D^{-\half}e^{-r(\ell\I+\la')^{-1}t}D^{\half}I_0\;.
\end{align*}
Note that $\la'$ is positive semi-definite and thus $\ell\I+\la'$ is invertible. Bearing in mind that $0$ is an eigenvalue of the matrix $\la'$ with the corresponding normalized eigenvector $\U_1=(\1^\top  D^{-1}\1)^{-\half}    D^{-\half}\1$, and by the spectral decomposition $\la'=\U\Lambda\U^\top $, we find that
\begin{align*}
I(t)&=D^{-\half}e^{-r\big(\U(\ell\I+\Lambda)\U^\top \big)^{-1}t}D^{\half}I_{0}
\\
&=D^{-\half}\U e^{-r(\ell\I+\Lambda)^{-1}t}\U^\top  D^{\half}I_{0}
\\
&=D^{-\half}\begin{bmatrix}
 \U_1 \;\; \tilde{\U}
\end{bmatrix} e^{-r
\begin{bmatrix}
\frac{1}{\ell} \;\;\;\;\;\;\;\; 0
\\
0_{ (n-1)\times 1} \; \tilde{\Lambda}
\end{bmatrix}
t}
\begin{bmatrix}
\U_1^\top 
\\
\tilde{\U}^\top 
\end{bmatrix}
D^{\half}I_{0}
\;,
\end{align*}
where $\tilde{\Lambda}={\rm diag}\{\frac{1}{\lambda_2+\ell},\;\frac{1}{\lambda_3+\ell},\cdots,\frac{1}{\lambda_n+\ell}\}$ and $0<\lambda_2<\lambda_3<\cdots<\lambda_n$ are nonzero eigenvalues of the matrix $\la'$. Let $\tilde{I}(t)=D^{\half}I(t)$. Noting that by the Kirchhoff Law, $\1^\top I_{0}=0$, we have
\begin{align*}
\tilde{I}(t)&=\tilde{\U} e^{-r\tilde{\Lambda}t}\;\tilde{\U}^\top  \tilde{I}_{0}\;.
\end{align*}
Since $\U^\top _1\tilde{I}_{0}=0$ we can write $\tilde{I}_{0}$ as the linear combination $\tilde{I}_0=\tilde{\U}X,\; X\in \R^{(n-1)\times 1}$.
Now we have
\begin{align*}
\tilde{I}(t)&=\tilde{\U} e^{-r\tilde{\Lambda}t}X, \qquad
%\end{align*}
%and
%\begin{align*}
\lVert \tilde{I}(t) \lVert^2=X^\top e^{-2r\tilde{\Lambda}t}X\;.
\end{align*}
Hence
\begin{align*}
\lVert \tilde{I}(t) \lVert^2 &\geq e^{-\frac{2r}{\lambda_2+\ell}t}\lVert \tilde{I}_{0} \lVert^2,\\
\quad
I^\top (t)DI(t) &\geq e^{-\frac{2r}{\lambda_2+\ell}t}I^\top _0DI_0,\\ \qquad \qquad
\lVert I(t) \lVert &\geq \mu e^{-\frac{r}{\lambda_2+\ell}t}\lVert I_0 \lVert,
\end{align*}
%\begin{align*}
%\lVert \tilde{I}(t) \lVert^2 &\geq e^{-\frac{2r}{\lambda_2+\ell}t}\lVert \tilde{I}_{0} \lVert^2
%\\
%I^\top (t)DI(t) &\geq e^{-\frac{2r}{\lambda_2+\ell}t}I^\top _0DI_0
%\\
%\lVert I(t) \lVert &\geq \mu e^{-\frac{r}{\lambda_2+\ell}t}\lVert I_0 \lVert\;,
%\end{align*}
where $$ \mu:=\sqrt{\frac{\min_i (\ell_{o_i})}{\max_i (\ell_{o_i})}}\;.$$
\\
This yields $\Psi_{\rm NIR}= \frac{\lambda_2+\ell}{r}$. Note that the eigenvalues of $\la'$ and $D\la$ are the same. This completes the proof.
\end{IEEEproof}
\begin{remark}\label{r:reviewer}
The results of Theorem \ref{t:LD} can be generalized to the case of non-uniform output impedances, each containing a nonzero resistor $r_o$ and a nonzero inductor $\ell_o$ in series. In this case, the network can be modeled by
\begin{align}\label{e:DrDl}
(r\I+\la D_r)I+(\ell\I+\la D_\ell) \dot{I}= \la V_o\;,
\end{align}
where $D_r:= \diag (r_{o_1},r_{o_2},\cdots,r_{o_n})$ and $D_\ell:= \diag (\ell_{o_1},\ell_{o_2},\cdots,\ell_{o_n})$.
% are assumed to be invertible, i.e. each output impedance contains a nonzero resistance and a nonzero inductance.
Analogously to the proof of Theorem \ref{t:LD}, it can be shown that the network inductivity ratio is calculated as
$$
	\Psi_{\rm NIR}=\min_{i\in\{2,3,\cdots,n\}} \frac{\lambda_{\ell_i}+\ell}{\lambda_{r_i}+r}\;\;,
$$
where $\lambda_{\ell_i}$ denotes the $i$th eigenvalue of the matrix $\la D_\ell $, and $\lambda_{\ell_1}=0$. Similarly, $\lambda_{r_i}$ denotes the $i$th eigenvalue of the matrix $\la D_r $, and $\lambda_{r_1}=0$.
\ER
\end{remark}

Exploiting the results of \cite{Merikoski2004}, bearing in mind that $\la$ and $D$ are both positive semi-definite matrices, we have
\begin{align}\label{LDbounds}
\lambda_2(\la) \min_i \ell_{o_i} \leq\lambda_2(D\la)\leq  \lambda_2(\la) \max_i \ell_{o_i}\;.
%\min_i (\ell_{o_i})\lambda_2(\la)\leq\lambda_2(D\la)\leq \max_i (\ell_{o_i})\lambda_2(\la)\;.
\end{align}
These bounds can be used to ensure that the network inductivity ratio lies within certain values, without explicitly calculating the algebraic connectivity of the directed graph associated with the Laplacian $D\la$.

{\em Optimizing the network inductivity ratio:}
The results proposed can be also exploited to maximize the diffusion of output impedances into the network.
This can be achieved by an optimal distribution of the inductors among the sources such that the network inductivity ratio is maximized. 
Below is an example illustrating this point on a network with a star topology. Another example using Kron reduction and phasors will be provided in Section \ref{s:examples}. 
%We assume that we have limited resources of inductors, which is reflected in the following {\em budget constraint}: 
%\begin{equation}\label{e:budget}
%\sum_{i}^{}\ell_{o_i}\leq c, \quad c\in\R^+.
%\end{equation}
%Note that the budget constraint can also be used to reflect any disadvantage resulting from a large output impedance, e.g. the voltage drop. According to Theorem \ref{t:LD}, this problem is equivalent to maximizing the algebraic connectivity of the directed graph $\hat{\mathcal{G}}$ under the budget constraint. The algebraic connectivity can be formulated as 
%%$$
%%\ell, \qquad l
%%$$
%\[
%\lambda_2(\hat{\mathcal{G}})=\min_{\upsilon \neq 0,\;  \upsilon^\top D^{-\half}\1=0}\;\;\frac{\upsilon^\top  D^{\frac{1}{2}}\la D^{\frac{1}{2}} \upsilon}{\lVert \upsilon  \lVert}.
%\]
%By defining $u:=D^{-\frac{1}{2}}\upsilon$, the optimization problem can be restated as follows:
%\begin{equation}\label{e:opt}
%\max_{D} \;\; 
%\min_{u \neq 0,\; u^\top  \1 =0}\;\;\frac{u^\top  D \la D u }{ \lVert D^{\frac{1}{2}} u  \lVert} \quad\quad \text{s.t.} \quad \text{trace}(D)\leq c\;.
%\end{equation}
%%Numerical methods can be used to obtain the optimal point of \eqref{e:opt}.
%Below, we provide an illustrative example where we solve the optimization problem above, and hence maximize the network inductivity ratio.
%
\begin{example}
	\begin{figure}
		\centering
		\includegraphics[width=8.8cm]{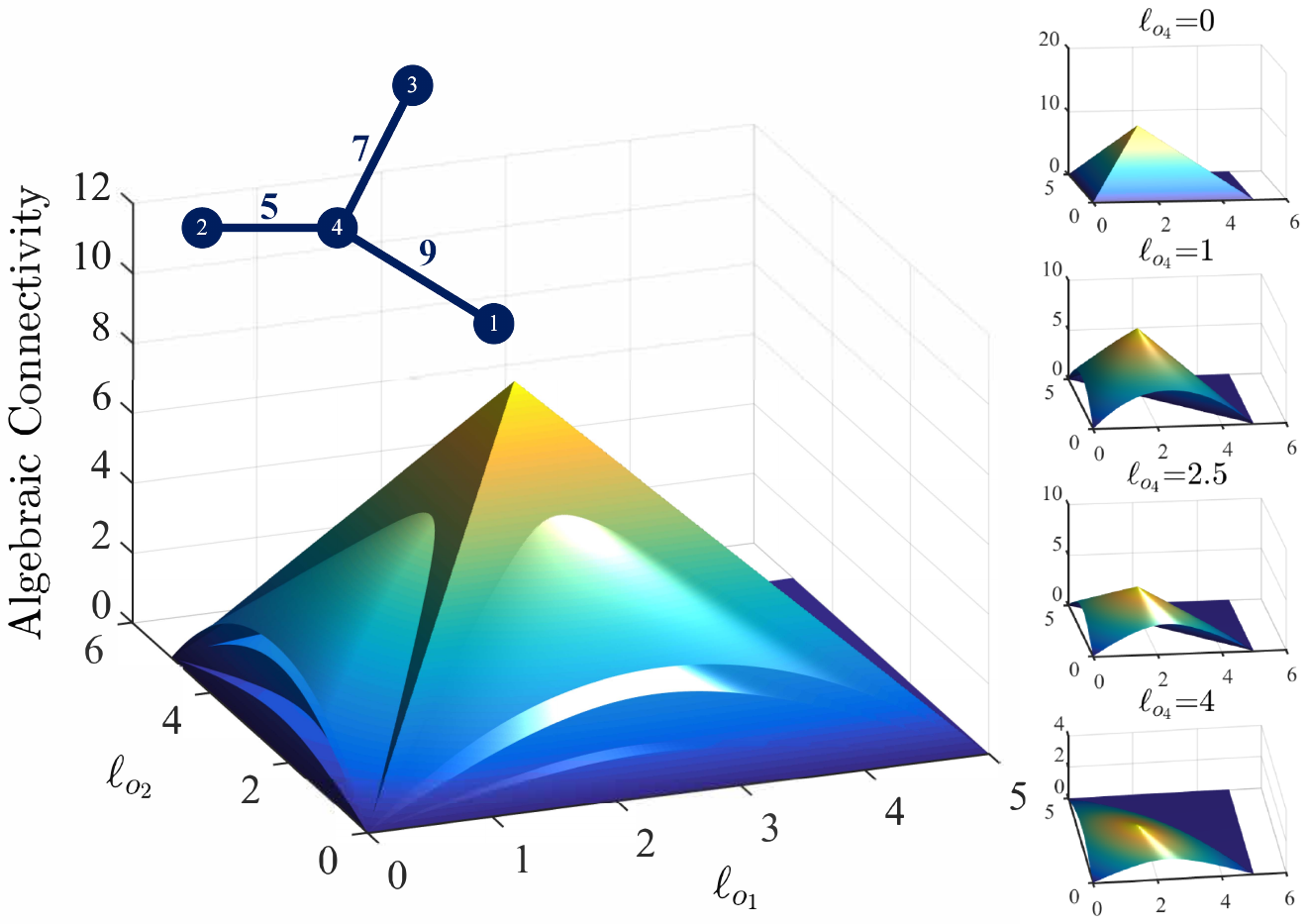}
		\caption{ Algebraic connectivity of the directed graph associated with the Laplacian $D\la$, where $\la$ is the Laplacian of the star graph shown in {top left corner}. The budget constraint is  $\sum_{i}\ell_{o_i}=c=5\mH$. On the right, the algebraic connectivity is plotted as a function of $\ell_{o_1}$ and $\ell_{o_2}$ for different values of $\ell_{o_4}$. Note that $\ell_{o_3}=c-\ell_{o_1}-\ell_{o_2}-\ell_{o_4}$. On the left, all the four subplots are merged. Clearly, the optimal algebraic connectivity is achieved where no output impedance is used for node $4$ (the middle node), and $\ell_{o_1}=2.20\mH$, $\ell_{o_2}=1.23\mH$, 
			$\ell_{o_3}=1.57\mH$ are used for nodes $1$, $2$, and $3$ respectively.}
		\label{fig:surf}
	\end{figure}
	%\textit{Output impedance optimization on a Star Graph.}\\
	Consider the graph $\mathcal{G}$ with the Laplacian $\la$, consisting of four nodes in a star topology. The line lengths are $5\pu$, $7\pu$, and $9\pu$, as depicted in Figure \ref{fig:surf}. Note that here again, the weights of the edges are the inverse of the distances. Attach the output impedances $D={\rm \diag}\{\ell_{o_1},\,\ell_{o_2},\,\ell_{o_3},\,\ell_{o_4}\}$ to each inverter, and assume that we have limited resources of inductors, namely \footnote{Note that the budget constraint \eqref{e:budget} can also be used to reflect any disadvantage resulting from a large output impedance, e.g. the voltage drop.}
\begin{equation}\label{e:budget}
\sum_{i}^{}\ell_{o_i}= c, \quad c\in\R^+.
\end{equation}
	%which is reflected in the following {\em budget constraint}: 
	%for simplicity, 
%assume that the budget constraint \eqref{e:budget} holds with equality. Hence $\ell_4=c-(\ell_1+\ell_2+\ell_3)$. 
Figure \ref{fig:surf} shows different values of the second smallest eigenvalue of the matrix $D\la$. We obtain that the maximal algebraic connectivity is achieved when no output impedance is used (wasted) for the node in the middle. 
Interestingly, the optimal value of output inductor for each node is proportional to its distance to the middle node.
\\
{Next we compare the performance of the droop controlled inverters in two different cases: i)  the network with the optimized output impedances as above, and ii) the network with evenly distributed output impedances. To this end, suppose that a source is connected to the middle node (node 4 in Figure \ref{fig:surf}), and three constant power loads are connected to the outer nodes (nodes 1, 2, and 3 in Figure \ref{fig:surf}) via droop-controlled power converters with the parameters given by Table \ref{tab:param}. Here, all the distribution lines are assumed to have the same reactance per length equal to $\w\ell=1.0\frac{\Omega}{\si{\kilo\meter}}$ and resistance per length equal to $r=0.1\frac{\Omega}{\si{\kilo\meter}}$.
%	 We compare the network with optimized output inductances with the case of evenly distributed one. 
	 At time $t=0$, the loads are increased with $10\%$ of their nominal value. The frequency of the inverters are shown in Figure \ref{fig:star3}.
	 It is evident that the droop controllers perform better in the network with the optimized network inductivity ratio $\Psi_{\rm NIR}$ (top), compared to the case of evenly distributed output inductances (below), where the solutions fail to converge.}
\ER
\begin{table}
	\centering
	{
	\caption{Simulation Parameters}}
	\label{tab:param}
	\begin{tabular}{ccccc}
		\toprule 
		& Load\textsubscript{1}  & Load\textsubscript{2}  & Load\textsubscript{3} & Source  \\
		\midrule
		Measurement Delay ($\si{\milli\second}$) &6  & 6 & 5 &8\\[0.15cm]
		Droop Coefficient (pu) & 0.07 & 0.14  & 0.14 & 0.28\\[0.15cm]
		Distance from the Source $(\si{\meter})$ &   90 &50&70&-\\[0.15cm]
		Voltage (pu) & 1.05& 1.10&0.95 &0.96\\[0.15cm]
		Nominal Active Power (pu)  & -0.9 & -0.8& -1.2 & 1.5 \\[0.1cm]
		\bottomrule
	\end{tabular}
	%	\vspace{-0.3cm}
\end{table}
\begin{figure}
\centering
\includegraphics[width=1\linewidth]{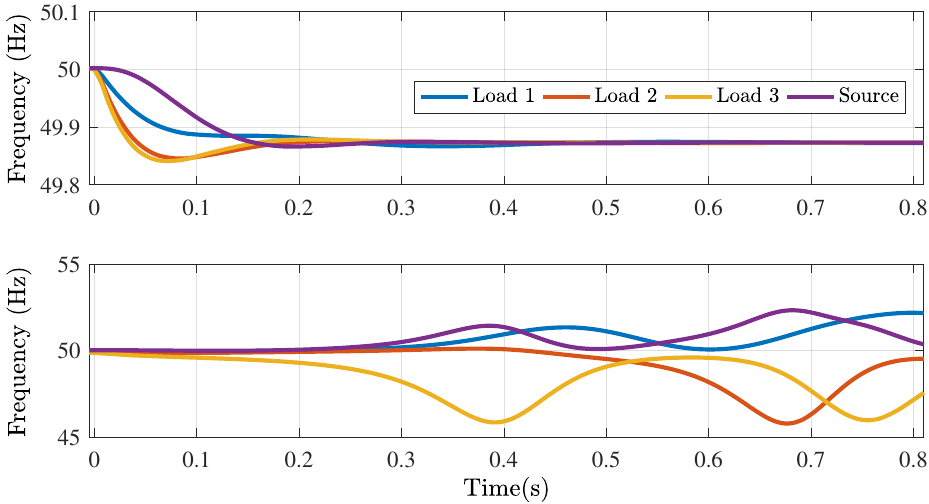}
\caption{{Comparison of droop controllers performance between the network with the optimized $\Psi_{\rm NIR}$ (top), and a network with evenly distributed output inductances (below).}}
\label{fig:star3}
\end{figure}
\end{example}
{
	The results of this subsection for non-uniform output impedances allow for the analysis of the networks containing non-tunable output impedances, e.g. constant impedance loads and synchronous generators/motors which can be modeled as a voltage source/sink behind a reactance. The optimization (and maximization of the algebraic connectivity) in this case involves only tuning the diagonal elements of the matrices $D_r$ and $D_l$ in \eqref{e:DrDl} associated with the tunable output impedances.
}

\section{Kron Reduction in Phasor Domain and the Network Inductivity Ratio}\label{s:examples}
\begin{figure}
\centering
\includegraphics[width=6.0cm]{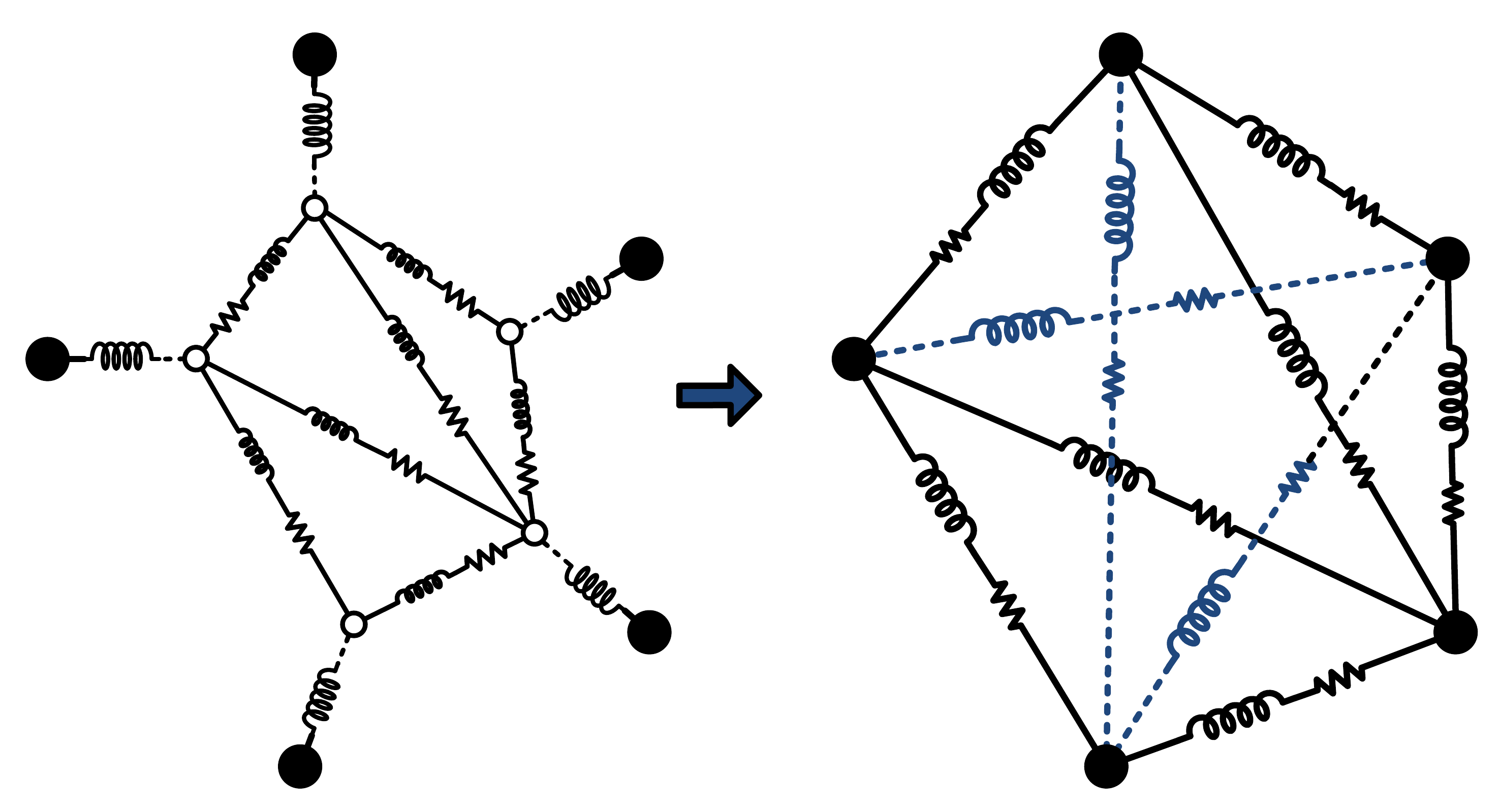}
\caption{Kron reduction of an arbitrary graph with added output impedances.}
\label{fig:kron}
\end{figure}

\begin{figure}
	\centering
	\includegraphics[width=8cm]{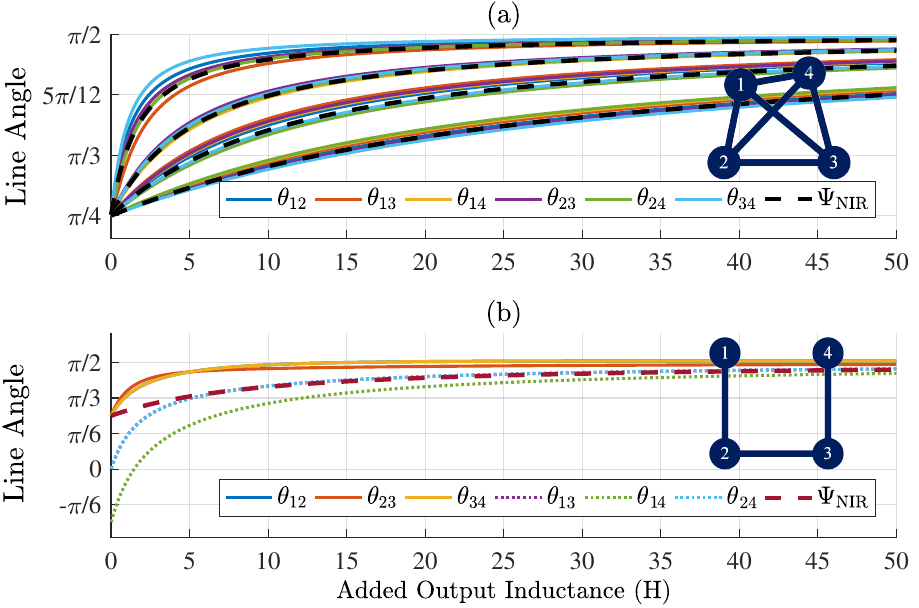}
	\caption{(a) Complete graph. The initial phase angle of the lines is $\frac{\pi}{4}$ and the line lengths are the following 4 cases (from top to bottom): [$\tau_{12},\,\tau_{13},\,\tau_{14},\,\tau_{23},\,\tau_{24},\,\tau_{34}$]$=$ [$5,\,6,\,9,\,7,\,4,\,6$]; [$20,\,19,\,20,\,21,\,20,\,22$]; [$40,\,37,\,35,\,49,\,46,\,38$]; [$100,\,105,\,93,\,87,\,110,\,89$]; (b) Path graph. Virtual lines are shown by dots. The initial phase angle of the lines is $\frac{\pi}{4}$ and the line lengths are equal to $5$.}
	\label{fig:completenonsymmetric}
\end{figure}

By leveraging Kron reduction and using the phasor domain, it is sometimes possible to synthesize an $RL$ circuit for the augmented network model \eqref{e:define} (see \cite{Willems2010} for more details). 
As depicted in Figure \ref{fig:kron}, in the Kron reduced graph some of the edges coincide with the lines of the original network into which the output impedances were diffused, while others (dotted {line}) are created as a result of the Kron reduction. We refer to the former as \textit{physical} and to the latter as \textit{virtual} lines.
%The Kron reduction does not necessarily preserve the homogeneity of the distribution lines, and hence except in some special cases, the equation \eqref{e:Laplacian} is not valid for the reduced network. %cannot be provided for such a network. 
%However, the reduced model can still be expressed by complex numbers in phasor domain, and the corresponding line angles can be used to validate our proposed inductivity metrics.  
%%bearing in mind that the output impedances are diffused into the lines of the reduced graph, to verify our method, we compare our proposed measure with the angles of the lines of the reduced graph in phasor domain. 
%To this end, 
To derive the Kron-reduced model, we first write the nodal currents as
\begin{align*}
\begin{bmatrix}
I
\\
0
\end{bmatrix}
=
\begin{bmatrix}
y_o\I && -y_o\I
\\
-y_o\I && y_o\I+ y_\ell \la
\end{bmatrix}
\begin{bmatrix}
V_{o}
\\
V
\end{bmatrix}\;,
\end{align*} 
where $$y_o=\frac{1}{j \w \ell_o},\;\;\; y_\ell=\frac{1}{r+j\w \ell}\;.$$ The Kron-reduced model is then obtained as
\begin{align}\label{e:red}
\Y_{\redu}=y_o[\I-(\I+\frac{y_\ell}{y_o}\la)^{-1}]\;.
\end{align}
Since every path between the outer nodes of the graph passes only through internal nodes (see Figure \ref{fig:kron}), the resulting Kron-reduced network is a complete graph \cite{Kron2013}. In the following example,
%Note that in some cases, the Kron reduction in Phasor domain leads to a description which cannot be realized with passive $RLC$ elements. 
%Here, however, 
%
%Below we provide a numerical example for the case of uniform output inductances added to the network. For the sake of clarity of the figure, 
we compare the line phase angles $$\theta_{ij}:=\arctan\frac{\rm{Im}(1/\Y_{\redu_{ij}})}{\rm{Re}(1/\Y_{\redu_{ij}})}$$ for the line $\{i,j\}$, to the phase angles suggested by $\Psi_{\rm NIR}$, namely 
\begin{align}\label{e:theta}
\theta_{\rm NIR}:=\arctan (\w\Psi_{\rm NIR})\;.
\end{align} 
Note that the term $\omega$ in the above is included to obtain reactance to resistance ratio from inductance to resistance ratio.
\begin{example}\label{ex:graph}
	\textit{Line angles of a Kron-reduced Graph with uniform output impedances.}\\
	\textbf{(a)} Consider a 4-node complete graph with different distribution line lengths. As shown in Figure \ref{fig:completenonsymmetric}a, the proposed measure matches with the overall behavior of the line angles as the added output inductance increases.
	\textbf{(b)} Consider a 4-node uniform path graph. Figure \ref{fig:completenonsymmetric}b shows that the least inductivity behavior is observed for 
	 the virtual lines. Hence the less virtual lines the reduced graph contains, the more output impedance diffuses in the network, which is consistent with our results in Section \ref{s:Calculating}. Also note that the inductance and resistance possess negative values at some edges for certain values of the output impedance. Therefore, it is difficult to extract a reasonable inductivity ratio for those edges from the Kron reduced phasor model. % fails to provide a meaningful inductivity ratio for those edges. 
 On the contrary, the proposed inductivity measure remains within the physically valid interval $\arctan (\omega\Psi_{\rm NIR})\in [0 \; \pi/2]$.
 \ER
\end{example}
Example \ref{ex:graph} shows that $\theta_{\rm NIR}$ can be used as a measure that estimates the phase of the lines of the overall network. A desired amount of change in this measure can be optimized by appropriate choices of output impedances. The following example illustrates this case.
\begin{figure}
	\centering
	\includegraphics[width=7cm]{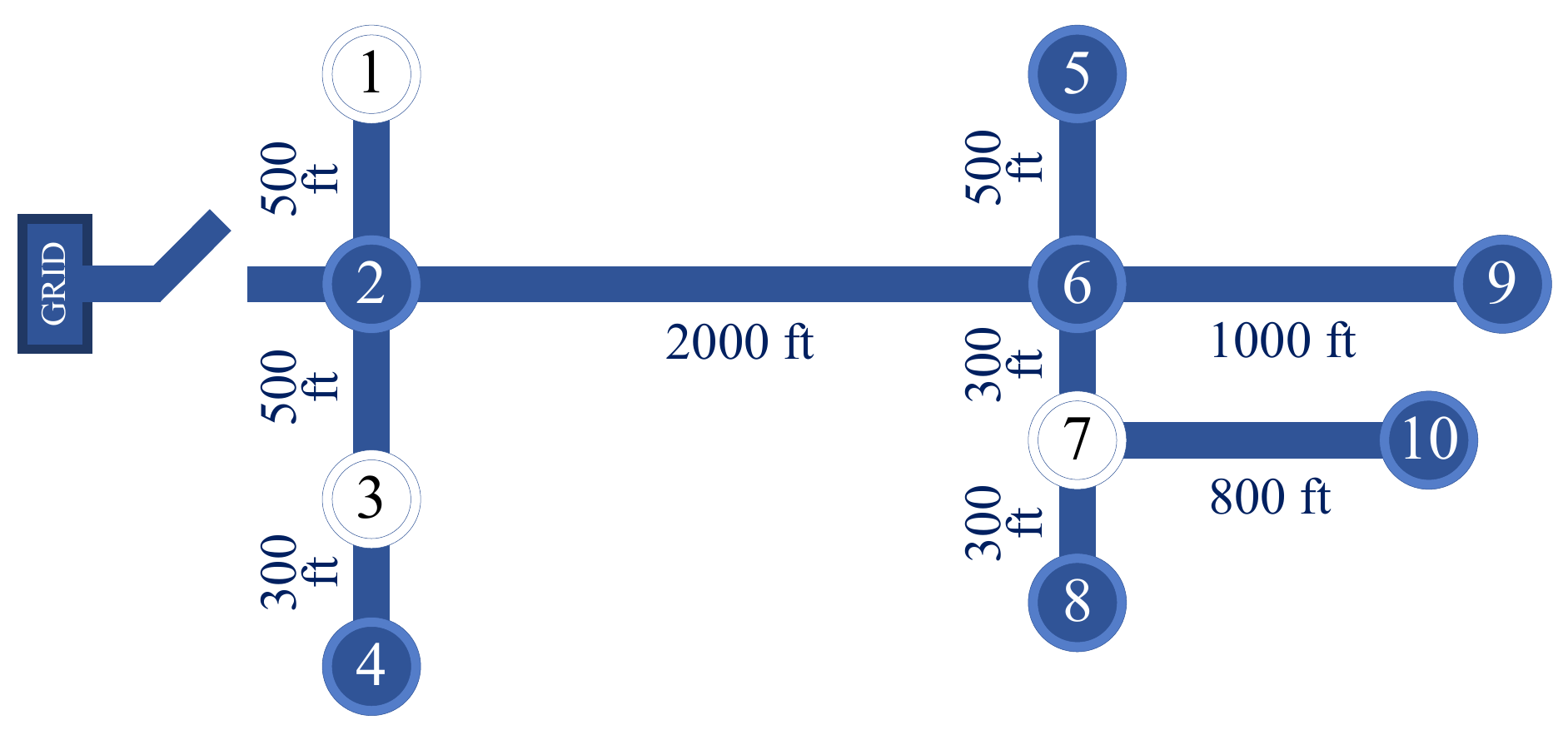}
	\caption{Schematic of the islanded IEEE 13 node test feeder graph. Power sources are connected to the ({white}) nodes $1$, $3$, and $7$.}
	\label{fig:IEEE13}
\end{figure}
\begin{example}
	\textit{Output impedance optimization on the IEEE 13 node test feeder.}\\
	Figure \ref{fig:IEEE13} depicts the graph of the islanded IEEE 13 node test feeder. Here, all the distribution lines are assumed to have the same reactance per length equal to $\w\ell=1.2 \frac{\Omega}{\text{mile}}$ and  resistance per length equal to $r=0.7 \frac{\Omega}{\text{mile}}$, derived from the configuration 602 \cite{IEEE13}. We consider the case where three inverters are connected to the nodes $1$, $3$, and $7$, and the rest of the nodes are connected to constant current loads. After carrying out Kron reduction, the algebraic connectivity of the resulting Laplacian matrix is equal to $3.1$. 
	%To be able to quantify the changes in line phase angles after addition of output impedances, we define $\theta_{\rm NIR}$ as
%\begin{align}\label{e:theta}
%\theta_{\rm NIR}:=\arctan (\w\Psi_{\rm NIR})\;.
%\end{align} 
Before adding the output inductances, $\theta_{\rm NIR}$ is equal to $arctan \frac{\w\ell}{r}=\frac{2\pi}{3}$. Based on the results of Theorem \ref{t:rl}, for a $10\%$ increase in $\theta_{\rm NIR}$, a uniform inductor of $3.21\mH$ should be attached to the outputs of all the sources. However, in case non-uniform output inductors are to be used, by using the result of Theorem \ref{t:LD}, the same increase in the network inductivity ratio can be achieved (optimally) with $0.95\mH$, $0.95\mH$, and $4.35\mH$ inductors, for the nodes $1$, $3$, and $7$ respectively. Both cases are feasible in practice since the typical values for inverter output filter inductance and implemented output virtual inductance range from $0.5\mH$ to $50\mH$ \cite{Li2009,Kim2011,Brabandere2007, V0, V1, V2, V3, droopfail3}. However, note that the total inductance used in the (optimal) non-uniform case is considerably smaller than the one used in the uniform scheme.
%Brabandere2007, , droopfail2, V0, V1, V2, V3, V4}.
%\vspace{0.1cm}
\ER
\end{example}
\section{Conclusion}\label{s:concl}
In this paper, the influence of the output impedance on the inductivity and resistivity of the distribution lines has been investigated. Two measures, network inductivity ratio and network resistivity ratio, were proposed and analyzed without relying on the ideal sinusoidal signals assumption (phasors). The analysis revealed the fact that the more connected the graph is, the more output impedance diffuses into the network and the larger its effect will be. We have provided examples on how the impact of inductive output impedances on the network can be maximized in specific network topologies. We compared the proposed measure to the phase angles of the lines in a phasor-based Kron reduced network. Results confirm the validity and the effectiveness of the proposed metrics. Future works include investigating analytical solutions on maximizing the network inductivity/resistivity ratios, {quantifying network inductivity in the case of heterogeneous lines, and investigating, analytically, the effect of network inductivity ratio on the performance of the droop-based methods.}  
\bibliographystyle{IEEETran}
\bibliography{ref}
\IEEEpeerreviewmaketitle
\end{document}